\documentclass[jmp, numerical]{revtex4}
\usepackage{array}
\usepackage{dcolumn}
\usepackage{lineno}
\usepackage{amsmath}
\usepackage{natbib}
\usepackage{hyperref}
\usepackage[all]{xy}

\newtheorem{proposition}{Proposition}

\newtheorem{example}[proposition]{Example}

\newtheorem{thm}[proposition]{Theorem}

\newtheorem{define}[proposition]{Definition}
\newtheorem{rem}[proposition]{Remark}

\newcommand{\R}{\mathbb{R}}
\newcommand{\C}{\mathbb{C}}

\makeatletter
\newcommand{\xyC}[1]{%
\makeatletter
\xydef@\xymatrixcolsep@{#1}
\makeatother
} 

\makeatletter
\newcommand{\xyR}[1]{%
\makeatletter
\xydef@\xymatrixrowsep@{#1}
\makeatother
} 

\begin{document}

\title{Gravitational Chern-Simons and the Adiabatic Limit}

\author{Brendan McLellan}
\address{Department of Mathematics, University of Toronto, Toronto,
Ontario, Canada M5S 2E4}
\email{mclellan@math.toronto.edu}



\begin{abstract}
We compute the gravitational Chern-Simons term explicitly for an adiabatic family of metrics using standard methods in general relativity.  We use the fact that our base three-manifold is a quasi-regular $K$-contact manifold heavily in this computation.  Our key observation is that this geometric assumption corresponds exactly to a Kaluza-Klein \emph{Ansatz} for the metric tensor on our three manifold, which allows us to translate our problem into the language of general relativity.  Similar computations have been performed in \cite{gijp}, although not in the adiabatic context.
\end{abstract}

\keywords{Contact manifold, Kaluza-Klein reduction}

\maketitle

\begin{acknowledgements}
The author would like to thank Roman Jackiw for generously taking the time to answer some questions pertaining to this paper and also for correcting some historical inaccuracies in an early draft.  He would also like to thank Lisa Jeffrey for several insightful discussions regarding the structure of this paper.
\end{acknowledgements}
\section{Introduction}
The gravitational Chern-Simons term was first introduced in the physics literature by S. Deser, R. Jackiw, G. 't Hooft and S. Templeton (See \cite{djt}, for example).  The reduction of the gravitational Chern-Simons term from three to two dimensions was subsequently investigated in \cite{gijp} for an abstract three-dimensional space-time.  A key observation in \cite{gijp} is that setting a Kaluza-Klein \emph{Ansatz} for the metric tensor effects a reduction from three to two dimensions for the gravitational Chern-Simons term.  Our first observation is that the natural associated metric for a quasi-regular K-contact three-manifold satisfies this Ansatz.  Although the work in \cite{gijp} necessarily computes the gravitational Chern-Simons term for our class of three-manifolds, we must extend the results of \cite{gijp} to perform this calculation for the \emph{adiabatic} family of metrics in Eq. \eqref{metfam}.\\
\\
The main goal of this paper is to study the gravitational Chern-Simons term with respect to an \emph{adiabatic} family of metrics.  We note that the terminology \emph{adiabatic limit} in this paper is meant to describe the limit in Eq. \eqref{gravlim} with respect to the \emph{adiabatic} family of metrics in Eq. \eqref{metfam}.  We borrow this terminology from \cite{bhr}, where one considers the \emph{adiabatic} limit for the family of eta-invariants,
\begin{equation}
\eta(\star_{\epsilon} d),
\end{equation}
where $d:\Omega^{1}(X)\rightarrow\Omega^{2}(X)$ is the standard exterior derivative on forms, $\star_{\epsilon}$ are the Hodge star operators for the family of metrics in Eq. \eqref{metfam}, and the eta-invariant is defined as usual,
\begin{equation}
\eta(\star d)(s):=\sum_{\lambda\in\text{spec}^{*}(\star d)}(sgn\lambda)|\lambda|^{-s},
\end{equation}
so that,
\begin{equation}
\eta(\star d):=\eta(\star d)(0).
\end{equation}
It is shown in \cite{bhr}, for example, that the adiabatic limit,
\begin{equation}
\lim_{\epsilon\rightarrow 0}\eta(\star_{\epsilon} d),
\end{equation}
exists, and in fact, for the case of quasi-regular K-contact three-manifolds, is a topological invariant of $X$.  As is noted in \cite{bhr}, the adiabatic limit has been known for some time, and has been studied in \cite{bc} and \cite{dai}, for example.\\
\\
This paper explicitly computes the \emph{adiabatic limit}
\begin{equation}\label{gravlim}
\lim_{\epsilon\rightarrow 0}\frac{\text{CS}(A^{g_{\epsilon}})}{2\pi},
\end{equation}
where,
\begin{equation}\label{gravchern}
\text{CS}(A^{g_{\epsilon}}):=\frac{1}{4\pi}\int_{X}Tr(A^{g_{\epsilon}}\wedge dA^{g_{\epsilon}}+\frac{2}{3} A^{g_{\epsilon}}\wedge A^{g_{\epsilon}}\wedge A^{g_{\epsilon}}),
\end{equation}
is the gravitational Chern-Simons term defined on a quasi-regular $K$-contact three-manifold, $(X,\phi,\xi,\kappa,g)$ (see \S\ref{geosec}).  Let $G=\operatorname{Spin}(6)$, $Q=TX\oplus TX$ viewed as a principal
$\operatorname{Spin}(6)$-bundle over $X$, $g\in\Gamma(S^{2}(T^{*}X))$ a Riemannian metric on $X$, $\phi:Q\rightarrow SO(X)$ a principal bundle morphism, and $A^{LC}\in\mathcal{A}_{SO(X)}:=\{ A\in (\Omega^{1}(SO(X))\otimes\frak{so}(3))^{SO(3)}\,\,|\,\,A(\xi^{\sharp})=\xi, \,\,\forall\,\xi\in\frak{so}(3)\}$ the Levi-Civita connection.  Then $A^{g}=\phi^{*}A^{LC}\in \mathcal{A}_{Q}:=\{ A\in (\Omega^{1}(Q)\otimes\frak{g})^{G}\,\,|\,\,A(\xi^{\sharp})=\xi, \,\,\forall\,\xi\in\frak{g}\}$.  We consider $A^{g_{\epsilon}}$ defined for the family of metrics,
\begin{equation}\label{metfam}
g_{\epsilon}=\epsilon\,\kappa\otimes\kappa+d\kappa(\cdot,J\cdot).
\end{equation}
where $\epsilon\in\mathbb{R}$ is a scalar parameter.\\
\\
The main motivation for the considerations of this paper come from previous work related to $U(1)$ Chern-Simons theory \cite{jm2}.  In this work, we are naturally led to compute the adiabatic limit of the regularized eta-invariant,
\begin{equation}\label{reg1}
\frac{\eta(\star_{\epsilon} d)}{4}+\frac{1}{12}\frac{\text{CS}(A^{g_{\epsilon}})}{2\pi}.
\end{equation}
for the case of a \emph{quasi-regular K-contact three-manifold} (see \S\ref{geosec}).  Equation \eqref{reg1} was introduced by Witten in \cite{w3} in order to cancel an anomaly that shows up in the computation of the stationary phase approximation for the Chern-Simons partition function.  By the Atiyah-Patodi-Singer theorem Eq. \eqref{reg1} is a topological invariant and our objective in \cite{jm2} was to compute this topological invariant explicitly.
\begin{rem}
Note that Eq. \eqref{reg1} actually depends on a choice of 2-framing for $X$, and by a result of Atiyah \cite{at} there exists a \emph{canonical} choice of such framing.  In our case, we explicitly choose a framing via a particular choice of Vielbein (See Eq.'s \eqref{vielchoice} and \eqref{vielchoice2}), and work with this throughout this paper.  In particular, we may view Eq. \eqref{reg1} as a topological invariant without ambiguity.
\end{rem}
It turns out that the computation of adiabatic limit of Eq. \eqref{reg1} is all that is required to obtain this explicit identification.  This fact, combined with our analysis in \cite{jm2}, led us to study the limit of Eq. \eqref{gravlim}.\\
\\
We note that our computations are modeled on \cite{gijp}.  The novelty in this article is the introduction of the adiabatic family of metrics in Eq. \eqref{metfam}, for which one must be careful to keep track of the $\epsilon$ dependence in the explicit computation of the gravitational Chern-Simons term of Eq. \eqref{gravchern}.\\
\\
The first thing we do is find an explicit formula for the family of metrics in Eq. \eqref{metfam} relative to a \emph{special coordinate system} that is adapted to our geometric situation.  We then observe that the result we obtain is precisely the Kaluza-Klein \emph{Ansatz} of \cite{gijp}.  This allows us to carry out the rest of our analysis out in parallel with \cite{gijp}.  We then compute the Christoffel symbols for the Levi-Civita connection for the family of metrics in Eq. \eqref{metfam}.  Using the formulae for the Christoffel symbols and Vielbein, we compute the components of the spin connection $A^{g_{\epsilon}}$ and directly evaluate $CS(A^{g_{\epsilon}})$. Our computation yields a formula for $CS(A^{g_{\epsilon}})$ in ``reduced'' terms as in \cite{gijp} and also provides an explicit identification of the $\epsilon$-dependence for this quantity.  We note that the class of quasi-regular K-contact three-manifolds are necessarily $U(1)$-bundles that fiber over an orbifold surface $\Sigma$.  Our main result is the following:
\begin{thm}\label{r5}
Let $(X,\phi,\xi,\kappa,g)$ be a closed, quasi-regular K-contact three-manifold,
\begin{displaymath}
\xymatrix{\xyC{2pc}\xyR{1pc}U(1) \ar@{^{(}->}[r] & X \ar[d]\\
                              & \Sigma}.
\end{displaymath}
Let $g_{\epsilon}:=\epsilon\,\kappa\otimes\kappa+\pi^{*}h$.  After a particular choice of Vielbein, (See Eq.'s \eqref{vielchoice} and \eqref{vielchoice2}) then,
\begin{equation}
CS(A^{g_{\epsilon}})=\left(\frac{\epsilon}{2}\right)\int_{\Sigma}r\,\omega+\left(\frac{\epsilon^{2}}{2}\right)\int_{\Sigma}f^{2}\,\omega
\end{equation}
where $r\in C^{\infty}_{orb}(\Sigma)$ is the (orbifold) scalar curvature of $(\Sigma,h)$, $\omega\in\Omega^{2}_{orb}(\Sigma)$ is the (orbifold) Hodge form of $(\Sigma,h)$, and $f\in\Omega^{0}_{orb}(\Sigma)$ is the invariant field strength on $(\Sigma,h)$.  In particular, the adiabatic limit of $\text{CS}(A^{g_{\epsilon}})$ vanishes:
\begin{equation}
\lim_{\epsilon\rightarrow 0}\text{CS}(A^{g_{\epsilon}})=0.
\end{equation}
\end{thm}


\section{Geometry}\label{geosec}
In this section we briefly review our geometric situation.  In particular, we recall the definition of a quasi-regular K-contact manifold and review some standard facts about these structures in the case of dimension three.
\begin{rem}
Our three-manifolds $X$ are assumed to be closed throughout this paper.
\end{rem}
\begin{define}\label{eq1}
A \emph{K-contact} manifold is a manifold $X$ with a contact metric structure $(\phi,\xi,\kappa,g)$ such that the Reeb field $\xi$ is Killing for the associated metric $g$, $\mathcal{L}_{\xi}g=0$.
\end{define}
\noindent
where,
\begin{itemize}
\item  $\kappa\in\Omega^{1}(X)$ contact form, $\xi\in \Gamma(TX)$ Reeb vector field.

\item  $H:=\text{ker}\kappa\subset TX$ denotes the horizontal or contact distribution on $(X,\kappa)$.

\item  $\phi\in \text{End}(TX)$, $\phi(Y)=JY$ for $Y\in \Gamma(H)$, $\phi(\xi)=0$ where $J\in \text{End}(H)$ complex structure on the contact distribution $H\subset TX$.

\item  $g=\kappa\otimes\kappa+d\kappa(\cdot, \phi\cdot)$

\end{itemize}
\noindent
\begin{rem}
Recall that the defining condition for a contact form $\kappa\in\Omega^{1}(X)$ is $\kappa\wedge d\kappa \neq 0$.  Also recall that the Reeb vector field $\xi\in \Gamma(TX)$ is uniquely defined by $\kappa$ by the conditions $\iota_{\xi}\kappa=1$ and $\iota_{\xi}d\kappa=0$, where $\iota_{\xi}$ denotes contraction of differential forms with the vector field $\xi$.  Note that we will assume that our contact structure is ``co-oriented,'' meaning that the contact form $\kappa\in\Omega^{1}(X)$ is a global form.  Generally, one can take the contact structure to be defined only locally by the condition $H:=\text{ker}\,\kappa$, where $\kappa\in\Omega^{1}(U)$ for open subsets $U\in X$ contained in an open cover of $X$.
\end{rem}
\begin{define}\label{eq2}
The characteristic foliation $\mathcal{F}_{\xi}$ of a contact manifold $(X,\kappa)$ is said to be \emph{quasi-regular} if there is a positive integer $j$ such that each point has a foliated coordinate chart $(U,x)$ such that each leaf of $\mathcal{F}_{\xi}$ passes through $U$ at most $j$ times.  If $j=1$ then the foliation is said to be \emph{regular}.
\end{define}
\noindent
Definitions \eqref{eq1} and \eqref{eq2} together define a quasi-regular $K$-contact manifold, $(X,\phi,\xi,\kappa,g)$.  Such three-manifolds are necessarily ``Seifert'' manifolds that fiber over a two dimensional orbifold $\Sigma$ with some additional structure.  Recall:
\begin{define}
A \emph{Seifert manifold} is a three-manifold $X$ that admits a locally free $U(1)$-action.
\end{define}
\noindent
Thus, Seifert manifolds are simply $U(1)$-bundles over an orbifold $\Sigma$,
\begin{displaymath}
\xymatrix{\xyC{2pc}\xyR{1pc}U(1) \ar@{^{(}->}[r] & X \ar[d]\\
                              & \Sigma}.
\end{displaymath}
\noindent
We have the following classification result:  $X$ is a quasi-regular K-contact three-manifold $\iff$
\begin{itemize}
\item  \cite[Theorem 7.5.1, (i)]{bg} $X$ is a $U(1)$-Seifert manifold over a Hodge orbifold surface, $\Sigma$.

\item  \cite[Theorem 7.5.1, (iii)]{bg} $X$ is a $U(1)$-Seifert manifold over a normal projective algebraic variety of real dimension two.
\end{itemize}
\begin{example}
All 3-dimensional lens spaces, $L(p,q)$ and the Hopf fibration $S^{1}\hookrightarrow S^{3}\rightarrow \C\mathbb{P}^{1}$ possess quasi-regular K-contact structures.  Note that any trivial $U(1)$-bundle over a Riemann surface $\Sigma_{g}$, $X=U(1)\times \Sigma_{g}$, possesses \emph{no} K-contact structure \cite{itoh}, however, and our results do not apply in this case.
\end{example}
\begin{rem}
Note that in fact our results apply to the class of all closed \emph{Sasakian} three-manifolds.  This follows from the observation that every Sasakian three-manifold is K-contact \cite[Corollary 6.5]{b}, and every K-contact manifold possesses a quasi-regular K-contact structure \cite[Theorem 7.1.10]{bg}.
\end{rem}
\noindent
A useful observation for us is that for a quasi-regular K-contact three-manifold, the metric tensor $g_{\epsilon}$ must take the following form \cite[Theorem 6.3.6]{bg}:
\begin{equation}
g_{\epsilon}=\epsilon\,\kappa\otimes\kappa+\pi^{*}h
\end{equation}
where $\pi:X\rightarrow \Sigma$ is our quotient map, and $h$ represents any (orbifold) K\"{a}hler metric on $\Sigma$ which is normalized so that the corresponding (orbifold) K\"{a}hler form, $\omega\in\Omega^{2}_{orb}(\Sigma,\R)$, pulls back to $d\kappa$.

\section{Gravitational Chern-Simons}
In order to perform the computation of the gravitational Chern-Simons term in Eq. \eqref{gravchern} explicitly, we will adopt the conventions of \cite{gijp}; indexing everything in sight and working in local coordinates.  To this end, we express the gravitational Chern-Simons term in a coordinate system $\{x^{0},x^{1},x^{2}\}$ as follows:
\begin{equation}\label{gravchernphys}
\text{CS}(A^{g_{\epsilon}}):=\frac{1}{4\pi}\int_{X}d^{3}x\,\,\epsilon^{\mu\nu\lambda}\,\,Tr\left((A^{g_{\epsilon}})_{\mu}\partial_{\nu} (A^{g_{\epsilon}})_{\lambda}+\frac{2}{3} (A^{g_{\epsilon}})_{\mu}(A^{g_{\epsilon}})_{\nu}(A^{g_{\epsilon}})_{\lambda}\right),
\end{equation}
where $\mu,\nu,\lambda\in\{0,1,2\}$, and $\epsilon^{\mu\nu\lambda}$ is the three-dimensional Levi-Civita symbol, normalized so that $\epsilon^{012}=1$, and defined for any $\sigma\in S_{3}=\{\text{permutations of}\,\, \{0,1,2\}\}$ by:
\begin{equation}
\epsilon^{\sigma(012)}:=(-1)^{|\sigma|},
\end{equation}
where $|\sigma|$ denotes the sign of sigma as a permutation, so that
\begin{equation}
|\sigma|:=\begin{cases} 0 & \text{if $\sigma$ is even,}
\\
1 &\text{if $\sigma$ is odd.}
\end{cases}
\end{equation}
Note that repeated indices are allowed, and our definition of $\epsilon^{\mu\nu\lambda}$ implies that $\epsilon^{\mu\nu\lambda}=0$ whenever any of $\mu,\nu,\lambda\in\{0,1,2\}$ are equal.\\
\\
A useful tool in our computation will be the \emph{Vielbein} formulation of gravity.  A \emph{Vielbein} is a local choice of orthogonal trivialization of the tangent bundle $TX$ of a (semi-) Riemannian manifold $(X,G)$, where $G$ denotes the metric tensor.
\begin{rem}  Note that we will generally write the notation $G$ for our metric interchangeably with $g_{\epsilon}$.  We will need to explicitly make the identification $G=g_{\epsilon}$ later, but for now this simplfies notation.  Also note that even though our choice of Vielbein in Eq.'s \eqref{vielchoice} and \eqref{vielchoice2} are \emph{locally} defined, it is shown in \cite[Eq. 3.30]{gijp} that this choice leads to a \emph{global} computation.  That is, the reduced gravitational Chern-Simons action is gauge and coordinate invariant, which is manifested precisely in the global expression of Eq. \eqref{finaleq}.
\end{rem}
Let us henceforth assume that $M=(X,G)$ is an oriented Riemannian three-manifold.  Given a local chart $U\subset X$, a Vielbein may therefore be expressed as a triple
\begin{equation}
\{\widetilde{E}_{0},\widetilde{E}_{1},\widetilde{E}_{2}\}
\end{equation}
where $\widetilde{E}_{A}\in\Gamma|_{U}(TX)$ such that
\begin{equation}\label{viel1}
G(\widetilde{E}_{A},\widetilde{E}_{B})=\eta_{AB}
\end{equation}
where $A,B\in\{0,1,2\}$.  Note that in a Lorentzian spacetime, $\eta_{AB}$ represents the Minkowski metric, of signature $(-,+,+)$ say, and in a Euclidean spacetime it represents the positive-definite Euclidean metric (so $\eta_{AB}=\delta_{AB}$, is the Kronecker pairing in this case).  We will work in a Euclidean signature in this article.  Given a choice of local coordinate system $\{x^{0},x^{1},x^{2}\}$ on $X$, we define
\begin{equation}
G_{\mu\nu}:=G(\partial_{\mu},\partial_{\nu})
\end{equation}
and we define the notation $\widetilde{E}^{\mu}_{A}$ to represent the coordinates of $\widetilde{E}_{A}$ in the coordinate system basis $\{\partial_{0},\partial_{1},\partial_{2}\}$:
\begin{equation}
\widetilde{E}_{A}=\sum_{\mu=0}^{2}\widetilde{E}^{\mu}_{A}\partial_{\mu}.
\end{equation}
Note that we also adopt the Einstein summation convention, and write
\begin{equation}
\widetilde{E}_{A}=\widetilde{E}^{\mu}_{A}\partial_{\mu},
\end{equation}
for example, where it is understood that a sum is taken over the repeated raised and lowered indices.
We may then express Eq. \eqref{viel1} in local coordinates:
\begin{equation}
G_{\mu\nu}\,\widetilde{E}^{\mu}_{A}\,\widetilde{E}^{\nu}_{B}=\eta_{AB}
\end{equation}
where $\mu,\nu\in\{0,1,2\}$ are thought of as indexing the spacetime coordinates related to the manifold coordinates $\{x^{0},x^{1},x^{2}\}$, and $A,B\in\{0,1,2\}$ are thought of as indexing the tangent space coordinates that label the Vielbein.  By dualizing to the cotangent bundle, we also consider
\begin{equation}
E^{A}:=E_{\mu}^{A}dx^{\mu}\in\Gamma|_{U}(T^{*}X).
\end{equation}
which are defined by requiring
\begin{equation}
E^{A}(\widetilde{E}_{B})=\delta^A_B.
\end{equation}
Let $G^{\mu\nu}$ denote the inverse of $G_{\mu\nu}$, so that $G^{\mu\lambda}\,G_{\lambda\nu}=G_{\nu\lambda}\,G^{\lambda\mu}=\delta^{\mu}_{\nu}$.
Our relevant relations for the Vielbein are then:
\begin{eqnarray}
\widetilde{E}^{\mu}_{A}E_{\nu}^{A}=\delta^{\mu}_{\nu}&,&\,\,\,\,\,\,\,\,E_{\mu}^{A}\widetilde{E}^{\mu}_{B}=\delta^{A}_{B}\\
G_{\mu\nu}\,\widetilde{E}^{\mu}_{A}\,\widetilde{E}^{\nu}_{B}=\eta_{AB}&,&\,\,\,\,\,\,\,\,G_{\mu\nu}=E_{\mu}^{A}E_{\nu}^{B}\eta_{AB}\\\label{vielrel}
E^{A}_{\mu}&=&G_{\mu\nu}\eta^{AB}\widetilde{E}^{\nu}_{B}.
\end{eqnarray}
Let $\nabla_{G}:\Gamma(TX)\rightarrow\Gamma(T^{*}X\otimes TX)$ denote the standard Levi-Civita connection associated to the metric $G$ on $X$.  Let
\begin{equation}
\Gamma^{\lambda}_{\mu\nu}:=\frac{1}{2}G^{\lambda\rho}\left(\partial_{\nu} G_{\rho\mu}+\partial_{\mu} G_{\rho\nu}-\partial_{\rho} G_{\mu\nu}\right)
\end{equation}
be the Christoffel symbols relative to the coordinate basis $\{x^{0},x^{1},x^{2}\}$ for the Levi-Civita connection, i.e.
\begin{equation}
(\nabla_{G})_{\partial_{\mu}}\partial_{\nu}=\Gamma^{\lambda}_{\mu\nu}\partial_{\lambda}.
\end{equation}

Our computation is facilitated by the basic relationship between the spin connection $A^{G}$, and the Levi-Civita connection $\nabla_{G}$ \cite[Eq. 2.17]{gijp}:
\begin{equation}\label{spinviel}
[(A^{G})_{\mu}]^{A}_{B}=E_{\nu}^{A}\,\widetilde{E}^{\lambda}_{B}\Gamma^{\nu}_{\mu\lambda}-\widetilde{E}^{\lambda}_{B}\partial_{\mu}E_{\lambda}^{A}.
\end{equation}
Our goal then is to compute the Christoffel symbols for the family of metrics $G=g_{\epsilon}$, which will give us an explicit formula for the spin connections $[(A^{G})_{\mu}]^{A}_{B}$.  We note that in three dimensions, the spin connection is anti-symmetric in $A, B$,
\begin{equation}
[(A^{G})_{\mu}]_{A\,B}:=\eta_{AC}[(A^{G})_{\mu}]^{C}_{B}=-[(A^{G})_{\mu}]_{B\,A},
\end{equation}
and we may use this fact to write
\begin{equation}\label{compeq}
[(A^{G})_{\mu}]_{A\,B}:=\eta_{AC}[(A^{G})_{\mu}]^{C}_{B}=\epsilon_{ABC}A^{C}_{\mu},
\end{equation}
where $A^{C}_{\mu}$ is a vector-valued one-form defined by this relation.  We then obtain a slightly simpler expression for $CS(A^{G})$ \cite[Eq. 2.22]{gijp}:
\begin{eqnarray}
\text{CS}(A^{g_{\epsilon}})&=&-\frac{1}{4\pi}\int_{X}d^{3}x\,\,\epsilon^{\mu\nu\lambda}\,\,\left(2\eta_{AB}A^{A}_{\mu}\partial_{\nu} A^{B}_{\lambda}-\frac{2}{3} \epsilon_{ABC}A^{A}_{\mu}A^{B}_{\nu}A^{C}_{\lambda}\right)\\\label{compeq2}
                           &=&-\frac{1}{2\pi}\int_{X}d^{3}x\,\,\epsilon^{\mu\nu\lambda}\,\,\left(\eta_{AB}A^{A}_{\mu}\partial_{\nu} A^{B}_{\lambda}\right)+\frac{1}{\pi}\int_{X}d^{3}x\,\,\text{det}A^{A}_{\mu}
\end{eqnarray}
Note that we have suppressed the $A^{g_{\epsilon}}$ notation in our integrals to just $A$.  Eq. \eqref{compeq} will be the result that we use to perform our computation directly.\\
\\
The first quantities that we wish to compute are the Christoffel symbols.  Before we can do this, however, we will need to find a useful expression for our family of metrics, $G=g_{\epsilon}$, in an ``arbitrary'' coordinate system $\{x^{0},x^{1},x^{2}\}$ on $X$ that reflects our geometric situation.  We follow \cite[pg. 265]{berg}, and introduce a ``special coordinate system.''  Such a coordinate system is adapted to our geometric situation in the following sense:  We define our coordinates via the local decomposition of $X$ given by a local trivialization from its bundle structure
\begin{equation}
\pi^{-1}(U)\simeq U\times S^{1},
\end{equation}
where $U\subset \Sigma$ is any open subset of $\Sigma$.
\begin{rem}
Note that $\Sigma$ is an orbifold.  All of our considerations are completely valid for the orbifold case by the results of Ichiro Satake \cite{sat}.  In particular, the notions of (co-) tangent bundles, (co-) tangent vectors, forms, curvature, integration, Riemannian metrics, orthogonal frames, etc... all have rigorously defined orbifold counterparts.  For example, the decomposition $\pi^{-1}(U)\simeq U\times S^{1}$ assumes that $U\subset\R^{2}$ is an \emph{orbifold} coordinate chart on $\Sigma$.  We do not wish to provide an in depth discussion of this here, and instead refer the reader to either \cite{sat}, or to \cite{nic} for a very readable account of these constructions.  For example, the decomposition $\pi^{-1}(U)\simeq U\times S^{1}$ above becomes $\pi^{-1}(\widehat{U})\simeq \widehat{U}\times S^{1}$ where $\widehat{U}\subset\R^{2}$ is an orbifold coordinate chart on $\Sigma$.
\end{rem}
\noindent
If $\xi$ denotes the Reeb vector field on our quasi-regular $K$-contact manifold, $(X,\phi,\xi,\kappa,g)$, then our coordinate system is chosen such that $\xi_{p}=[0,0,1]$, for any $p\in U$, and the first two coordinates $\{x^{0},x^{1}\}$ coincide with the coordinates on our base manifold $\Sigma$.  We should note that such a choice of coordinates does not necessarily respect the contact structure, $TX \simeq H\oplus \R \xi$, of our K-contact manifold $(X,\phi,\xi,\kappa,g)$;  i.e. for the coordinates $\{x^{0},x^{1}\}$, the associated vector fields $\frac{\partial}{\partial x^{0}},\frac{\partial}{\partial x^{1}}$ are not necessarily \emph{horizontal} vector fields.  The vector fields $\frac{\partial}{\partial x^{0}},\frac{\partial}{\partial x^{1}}$ may have components in the Reeb direction:
\begin{equation}
\frac{\partial}{\partial x^{\alpha}}=\textbf{h}_{\alpha}+\varphi_{\alpha}\frac{\partial}{\partial x^{2}},\,\,\,\alpha\in\{0,1\},
\end{equation}
where $\textbf{h}_{\alpha}$ is the horizontal component of $\frac{\partial}{\partial x^{\alpha}}$, and $\varphi_{\alpha}\frac{\partial}{\partial x^{2}}$ is its vertical component in the direction of the Reeb field.  Clearly, we have chosen our coordinates so that the local vector field $\frac{\partial}{\partial x^{2}}$ coincides with the Reeb direction.  We now wish to express our family of metrics in this coordinate system.  By definition:
\begin{equation}
g_{\epsilon}=\epsilon\,\kappa\otimes\kappa+\pi^{*}h.
\end{equation}
Evaluating this in our coordinate system yields:
\begin{equation}\label{glocmat}
G_{\mu\nu}=g_{\epsilon}=\begin{bmatrix}
h_{\alpha\beta}+\epsilon\varphi_{\alpha}\varphi_{\beta} & \epsilon\varphi_{\alpha} \\
\epsilon\varphi_{\beta}  & \epsilon
\end{bmatrix}.
\end{equation}
Our matrix is indexed with the understanding that $\alpha,\beta\in\{0,1\}$ and $\mu, \nu\in\{0,1,2\}$ index the entire matrix.  We follow \cite{gijp} in our notation, and letters from the middle Greek alphabet $(\lambda, \mu,\nu,\ldots)$ will denote spacetime components on our three-manifold, while beginning Greek letters $(\alpha, \beta, \gamma,\ldots)$ will denote spacetime components on our reduced two-manifold.  Tangent space components are generally described by Latin letters, upper case $(A,B,C,\ldots)$ for three-dimensions and lower case $(a,b,c,\ldots)$ for two-dimensions.
\begin{rem}  Note that the $K$-contact condition (i.e. the Reeb field is Killing for the metric $G=g_{\epsilon}$) is crucial for our analysis and ensures that the quantities, $h_{\alpha\beta}, \varphi_{\alpha}$, are independent of the third local coordinate $x^{2}$, \cite[Eq. 17.60]{berg}.  It is precisely this condition that makes our computation of the gravitational Chern-Simons term feasible.  This result is easy to see in our special coordinate system $\{x^{0},x^{1},x^{2}\}$ described above.  Recall the definition of the Lie derivative,
\begin{equation}
\mathcal{L}_{\xi}g:=\frac{d}{dt}\Big|_{t=0}\phi^{*}_{t}g,
\end{equation}
where $\phi_{t}:X\rightarrow X$ is the flow of $\xi$.  In our coordinate system, $\phi_{t}(x^{0},x^{1},x^{2})=(x^{0},x^{1},x^{2}+t)$ and the condition $\frac{d}{dt}\Big|_{t=0}\phi^{*}_{t}g=0$ is equivalent to $\frac{d}{dt}\Big|_{t=0}g\circ\phi_{t}=\frac{d}{dt}\Big|_{t=0}g(x^{0},x^{1},x^{2}+t)=0$.    Using our explicit expression for the matrix $g$ in Eq. \eqref{glocmat} above we have
\begin{equation}\label{eqnphiz}
\frac{d}{dt}\Big|_{t=0}\varphi_{\alpha}(x^{0},x^{1},x^{2}+t)=\partial_{2}\varphi_{\alpha}(x^{0},x^{1},x^{2})=0,
\end{equation}
and,
\begin{eqnarray*}
\frac{d}{dt}\Big|_{t=0}(h_{\alpha\beta}+\varphi_{\alpha}\varphi_{\beta})(x^{0},x^{1},x^{2}+t)&=&\partial_{2}h_{\alpha\beta}+\partial_{2}(\varphi_{\alpha}\varphi_{\beta})\\
                                                                                             &=&\partial_{2}h_{\alpha\beta},\,\,\text{since $\partial_{2}\varphi_{\alpha}(x^{0},x^{1},x^{2})=0$ by Eq. \eqref{eqnphiz},}\\
                                                                                             &=&0.
\end{eqnarray*}
Thus, $\partial_{2}\varphi_{\alpha}=0$ and $\partial_{2}h_{\alpha\beta}=0$.
\end{rem}
We can now see that the Kaluza-Klein Ansatz of \cite[Eq. 3.28]{gijp} for the metric tensor is given by Eq. \eqref{glocmat} and is implied by our geometric situation (i.e. the assumption that $(X,\phi,\xi,\kappa,g)$ is a closed, quasi-regular K-contact three-manifold).  Note that our sign conventions differ, and we follow \cite[Eq. 17.53]{berg}.  Also, it is not difficult to show that our inverse metric is given by:
\begin{equation}
G^{\mu\nu}=\begin{bmatrix}
h^{\alpha\beta} & -h^{\alpha \delta}\varphi_{\delta} \\
-h^{\beta \delta}\varphi_{\delta}   & \epsilon^{-1}+h^{\delta \zeta}\varphi_{\delta}\varphi_{\zeta}
\end{bmatrix},
\end{equation}
\noindent
where $h^{\alpha\beta}$ denotes the inverse of the (orbifold) K\"{a}hler metric $h$ on $\Sigma$.  After some calculation, we find that the Christoffel symbols
\begin{equation}
\Gamma^{\lambda}_{\mu\nu}=\frac{1}{2}G^{\lambda\rho}\left(\partial_{\nu} G_{\rho\mu}+\partial_{\mu} G_{\rho\nu}-\partial_{\rho} G_{\mu\nu}\right),
\end{equation}
\noindent
for the metric $G=g_{\epsilon}$ may be computed as (see \ref{civita}):
\begin{eqnarray}
\Gamma^{\delta}_{\alpha\beta}&=&\gamma^{\delta}_{\alpha\beta}-\frac{\epsilon}{2}h^{\delta \zeta}(\varphi_{\beta}f_{\zeta\alpha}+\varphi_{\alpha}f_{\zeta\beta})\\
\Gamma^{2}_{\alpha\beta}&=&\frac{1}{2}(D_{\alpha}\varphi_{\beta}+D_{\beta}\varphi_{\alpha})+\frac{\epsilon}{2}\varphi^{\zeta}(\varphi_{\beta}f_{\zeta\alpha}+\varphi_{\alpha}f_{\zeta\beta}),
\\
\Gamma^{\delta}_{2\beta}&=&\frac{\epsilon}{2}h^{\delta\zeta}f_{\beta\zeta}\\
\Gamma^{2}_{2\beta}&=&\frac{\epsilon}{2}\varphi^{\zeta}f_{\zeta\beta}\\
\Gamma^{\delta}_{22}&=&\Gamma^{2}_{22}=0.
\end{eqnarray}
\noindent
Some explanation of notation is in order.  First, $\alpha, \beta, \delta, \zeta\in\{0,1\}$ index the coordinates on $\Sigma$.  The $\gamma^{\delta}_{\alpha\beta}$ are the Christoffel symbols for the Levi-Civita connection of the metric $h$ on $\Sigma$:
\begin{equation}
\gamma^{\delta}_{\alpha\beta}:=\frac{1}{2}h^{\delta\zeta}\left(\partial_{\beta} h_{\zeta\alpha}+\partial_{\alpha} h_{\zeta\beta}-\partial_{\zeta} h_{\alpha\beta}\right).
\end{equation}
\noindent
$D$ is the covariant derivative for the Levi-Civita connection of the metric $h$ on $\Sigma$:
\begin{equation}
D_{\alpha}\varphi_{\beta}:=\partial_{\alpha}\varphi_{\beta}-\gamma^{\zeta}_{\alpha\beta}\varphi_{\zeta}.
\end{equation}
\noindent
$f_{\alpha \beta}$ is the ``abelian field strength'' tensor:
\begin{equation}
f_{\alpha \beta}:=\partial_{\alpha}\varphi_{\beta}-\partial_{\beta}\varphi_{\alpha},
\end{equation}
\noindent
and finally, all two-dimensional indices are raised and lowered with the metric $h$; i.e.
\begin{equation}
\varphi^{\alpha}:=h^{\alpha\zeta}\varphi_{\zeta}.
\end{equation}
\noindent
In order to compute our spin connection using Eq. \eqref{spinviel}, we need an explicit formula for the Vielbein.  We choose these as follows:
\begin{eqnarray}\label{vielchoice}
E^{a}_{\alpha}&=&e^{a}_{\alpha},\,\,E^{2}_{2}=\sqrt{\epsilon},\,\,E^{2}_{\alpha}=\sqrt{\epsilon}\varphi_{\alpha},\,\,E^{a}_{2}=0\\\label{vielchoice2}
\widetilde{E}^{\alpha}_{a}&=&\widetilde{e}^{\alpha}_{a},\,\,\widetilde{E}^{2}_{2}=\frac{1}{\sqrt{\epsilon}},\,\,\widetilde{E}^{2}_{a}=-\varphi_{\zeta}\widetilde{e}^{\zeta}_{a},\,\,\widetilde{E}^{\alpha}_{2}=0,
\end{eqnarray}
\noindent
where $e^{a}_{\alpha}, \widetilde{e}^{\alpha}_{a}$ are the Vielbein (i.e. \emph{Zweibein}) for the two-dimensional metric tensor $h$ on $\Sigma$.  Note that $a, \alpha, \zeta\in\{0,1\}$ are indices in two-dimensions, and $a\in\{0,1\}$ denotes the ``tangent space coordinates'' that index the Zweibein, as usual.  We leave the straightforward confirmation that these formulae define a local orthogonal trivialization to the reader.  Thus, using Eq. \eqref{spinviel},
\begin{equation}
[A_{\mu}]^{A}_{B}=E_{\nu}^{A}\,\widetilde{E}^{\lambda}_{B}\Gamma^{\nu}_{\mu\lambda}-\widetilde{E}^{\lambda}_{B}\partial_{\mu}E_{\lambda}^{A},
\end{equation}
\noindent
and our formulae for the Vielbein and the connection components, we may compute the spin connections $[(A^{G})_{\mu}]^{A}_{B}:=[A_{\mu}]^{A}_{B}$ (see \ref{spin}):
\begin{eqnarray}
[A_{\alpha}]^{a}_{b}&=&\widetilde{e}^{\zeta}_{b}\left[-D_{\alpha}e^{a}_{\zeta}-\frac{\epsilon}{2}e^{a}_{\delta}h^{\delta\rho}\varphi_{\alpha}f_{\rho\zeta}\right]\\
\,[A_{\alpha}]_{a\,2}&=&-[A_{\alpha}]_{2\,a}=\widetilde{\eta}_{a\,b}\frac{\sqrt{\epsilon}}{2}e^{b}_{\delta}h^{\delta\zeta}f_{\alpha\zeta}\\
\,[A_{2}]^{a}_{b}&=&\frac{\epsilon}{2}\widetilde{e}^{\zeta}_{b}e^{a}_{\delta}h^{\delta\rho}f_{\zeta\rho}\\
\,[A_{2}]_{2\,a}&=&[A_{2}]_{a\,2}=0\\
\,[A_{\alpha}]^{2}_{2}&=&[A_{2}]^{2}_{2}=0
\end{eqnarray}
\noindent
where we have lowered the two dimensional index on $[A_{\alpha}]_{a\,2}:=\widetilde{\eta}_{ab}[A_{\alpha}]^{b}_{2}$ above.  $\widetilde{\eta}_{ab}$ is the two-dimensional Kronecker pairing.  Then using Eq. \eqref{compeq} we may compute the quantities $A^{C}_{\mu}$ as follows (see \ref{spinred}):
\begin{eqnarray}\label{spin1}
A^{2}_{\alpha}=-\omega_{\alpha}-\frac{\epsilon}{2}f\varphi_{\alpha}&,&\,\,\,A^{2}_{2}=-\frac{\epsilon}{2}f\\\label{spin2}
A^{a}_{\alpha}=\frac{\sqrt{\epsilon}}{2}e^{a}_{\alpha}f&,&\,\,\,A^{a}_{2}=0
\end{eqnarray}
\noindent
where $\omega_{\alpha}$ is defined by the relation $\eta_{ac}(\omega_{\alpha})^{c}_{b}=:\omega_{\alpha,ab}=\epsilon_{ab}\omega_{\alpha}$, and $(\omega_{\alpha})^{a}_{b}$ is the spin connection on $\Sigma$:
\begin{eqnarray}
(\omega_{\alpha})^{a}_{b}&:=&\widetilde{e}^{\zeta}_{b}\partial_{\alpha}e^{a}_{\zeta}-e^{a}_{\delta}\widetilde{e}^{\zeta}_{b}\gamma^{\delta}_{\alpha \zeta}\\
                         &=&\widetilde{e}^{\zeta}_{b}D_{\alpha}e^{a}_{\zeta}
\end{eqnarray}
Also, $f\in C^{\infty}_{orb}(\Sigma)$ is the \emph{invariant field strength} defined by the relation:
\begin{equation}
f_{\alpha\beta}=\sqrt{h}\,\epsilon_{\alpha\beta}\,f.
\end{equation}
Thus, using Eq.'s \eqref{spin1} and \eqref{spin2} and the formula for $CS(A^{g_{\epsilon}})$ given by Eq. \eqref{compeq2}, we find that (see \ref{reduced}):
\begin{eqnarray}
CS(A^{g_{\epsilon}})&=&-\frac{1}{4\pi}\int_{S^{1}}dx^{2}\int_{\Sigma}dx^{0}\wedge dx^{1}\sqrt{h}(\epsilon f r+\epsilon^{2} f^{3})\\\label{lastcal}
                    &=&-\frac{1}{2}\int_{\Sigma}dx^{0}\wedge dx^{1}\sqrt{h}(\epsilon f r+\epsilon^{2} f^{3})
\end{eqnarray}
where we take the volume of $S^{1}$ to be $2\pi$, $r\in \Omega^{0}_{orb}(\Sigma)$ is the (orbifold) scalar curvature, and
$\omega\in\Omega^{2}_{orb}(\Sigma)$ is the (orbifold) K\"{a}hler form of $\Sigma$.  Recall that the Reeb vector field is dual to the one-form $\kappa$ under our metric $g_{\epsilon}$ when $\epsilon=1$:
\begin{equation}
\kappa(\cdot)=g_{1}(\xi,\cdot).
\end{equation}
In our coordinate system, this means:
\begin{equation}
\kappa=\varphi_{0}dx^{0}+\varphi_{1}dx^{1}+dx^{2}.
\end{equation}
We then have:
\begin{eqnarray}
d\kappa=(\partial_{0}\varphi_{1}-\partial_{1}\varphi_{0})dx^{0}\wedge dx^{1}&=&f_{01}dx^{0}\wedge dx^{1},\\
                                                                            &=&\sqrt{h} f dx^{0}\wedge dx^{1}
\end{eqnarray}
We have implicitly identified the (orbifold) K\"{a}hler form $\omega$ on $\Sigma$ with its pullback under $\pi:X\rightarrow \Sigma$ here in this coordinate system.  Strictly speaking we have:
\begin{equation}
d\kappa=\pi^{*}\omega.
\end{equation}
By reversing the orientation of $\Sigma$, and substituting $\omega$ in for $\sqrt{h} f dx^{0}\wedge dx^{1}$ in Eq. \eqref{lastcal}, we obtain:
\begin{thm}
Let $(X,\phi,\xi,\kappa,g)$ be a closed, quasi-regular K-contact three-manifold,
\begin{displaymath}
\xymatrix{\xyC{2pc}\xyR{1pc}U(1) \ar@{^{(}->}[r] & X \ar[d]\\
                              & \Sigma}.
\end{displaymath}
Let $g_{\epsilon}:=\epsilon\,\kappa\otimes\kappa+\pi^{*}h$.  After a particular choice of Vielbein (See Eq's \eqref{vielchoice} and \eqref{vielchoice2}.) then,
\begin{equation}\label{finaleq}
CS(A^{g_{\epsilon}})=\left(\frac{\epsilon}{2}\right)\int_{\Sigma}r\,\omega+\left(\frac{\epsilon^{2}}{2}\right)\int_{\Sigma}f^{2}\,\omega
\end{equation}
where $r\in C^{\infty}_{orb}(\Sigma)$ is the (orbifold) scalar curvature of $(\Sigma,h)$, $\omega\in\Omega^{2}_{orb}(\Sigma)$ is the (orbifold) Hodge form of $(\Sigma,h)$, and $f\in\Omega^{0}_{orb}(\Sigma)$ is the invariant field strength on $(\Sigma,h)$.  In particular, the adiabatic limit of $\text{CS}(A^{g_{\epsilon}})$ vanishes:
\begin{equation}
\lim_{\epsilon\rightarrow 0}\text{CS}(A^{g_{\epsilon}})=0.
\end{equation}
\end{thm}
\appendix
\section{Computations}

\subsection{Levi-Civita Connection}\label{civita}
In this section we explicitly compute the Christoffel symbols
\begin{equation}
\Gamma^{\lambda}_{\mu\nu}=\frac{1}{2}G^{\lambda\rho}\left(\partial_{\nu} G_{\rho\mu}+\partial_{\mu} G_{\rho\nu}-\partial_{\rho} G_{\mu\nu}\right),
\end{equation}
for the Levi-Civita connection $\nabla_{G}$ for the family of metrics
\begin{equation}\label{metric1}
G_{\mu\nu}:=g_{\epsilon}=\begin{bmatrix}
h_{\alpha\beta}+\epsilon\varphi_{\alpha}\varphi_{\beta} & \epsilon\varphi_{\alpha} \\
\epsilon\varphi_{\beta}  & \epsilon
\end{bmatrix}.
\end{equation}
\noindent
with inverse metric
\begin{equation}\label{metric2}
G^{\mu\nu}=\begin{bmatrix}
h^{\alpha\beta} & -\varphi^{\alpha} \\
-\varphi^{\beta}   & \epsilon^{-1}+\varphi^{\zeta}\varphi_{\zeta}
\end{bmatrix}.
\end{equation}
\begin{rem}
We will use the comma notation to denote partial derivatives; i.e. $G_{\rho\beta,\alpha}:=\partial_{\alpha} G_{\rho\beta}$, and the semi-colon notation to denote covariant derivatives, so $\omega_{\alpha\beta;\rho}:=\partial_{\rho}\omega_{\alpha\beta}-\Gamma^{\zeta}_{\alpha\rho}\omega_{\zeta\beta}-\Gamma^{\zeta}_{\rho\beta}\omega_{\alpha\zeta}$ for some $(0,2)$ tensor $\omega_{\alpha\beta}$, for example.
\end{rem}
We break this computation down into cases.
\begin{enumerate}\renewcommand{\theenumi}{\Roman{enumi}}
\item  $\Gamma^{\delta}_{\alpha\beta}$, $\delta, \alpha, \beta\in\{0,1\}$.  Reading off the components of the metric from Eq.'s \ref{metric1} and \ref{metric2}:
\begin{itemize}
\item  $\rho\in\{0,1\}$:
\begin{eqnarray*}
&&\frac{1}{2}G^{\delta\rho}\left(G_{\rho\beta,\alpha}+G_{\rho\alpha,\beta}- G_{\alpha\beta,\rho}\right)\\
                             &=&\frac{1}{2}h^{\delta\rho}\left([h_{\rho\beta,\alpha}+\epsilon(\varphi_{\rho}\varphi_{\beta})_{,\alpha}]+[h_{\rho\alpha,\beta}+\epsilon(\varphi_{\rho}\varphi_{\alpha})_{,\beta}]- [h_{\alpha\beta,\rho}+\epsilon(\varphi_{\alpha}\varphi_{\beta})_{,\rho}]\right)\\
                             &=&\frac{1}{2}h^{\delta\rho}\left([h_{\rho\beta,\alpha}+\epsilon(\varphi_{\rho,\alpha}\varphi_{\beta}+\varphi_{\rho}\varphi_{\beta,\alpha})]+[h_{\rho\alpha,\beta}+\epsilon(\varphi_{\rho,\beta}\varphi_{\alpha}+\varphi_{\rho}\varphi_{\alpha,\beta})]\right)+\\
                             &-&\frac{1}{2}h^{\delta\rho}\left([h_{\alpha\beta,\rho}+\epsilon(\varphi_{\alpha,\rho}\varphi_{\beta}+\varphi_{\alpha}\varphi_{\beta,\rho})]\right)\\
                             &=&\gamma^{\delta}_{\alpha\beta}+\frac{\epsilon}{2}h^{\delta\rho}\left((\varphi_{\rho,\alpha}\varphi_{\beta}+\varphi_{\rho}\varphi_{\beta,\alpha})+(\varphi_{\rho,\beta}\varphi_{\alpha}+\varphi_{\rho}\varphi_{\alpha,\beta})-(\varphi_{\alpha,\rho}\varphi_{\beta}+\varphi_{\alpha}\varphi_{\beta,\rho})\right)\\
                             &=&\gamma^{\delta}_{\alpha\beta}+\frac{\epsilon}{2}\left(h^{\delta\rho}[\varphi_{\beta}(\varphi_{\rho,\alpha}-\varphi_{\alpha,\rho})+\varphi_{\alpha}(\varphi_{\rho,\beta}-\varphi_{\beta,\rho})]+\left\{\varphi^{\delta}\varphi_{\alpha,\beta}+\varphi^{\delta}\varphi_{\beta,\alpha}\right\}\right)
\end{eqnarray*}
where
\begin{equation}
\gamma^{\delta}_{\alpha\beta}:=\frac{1}{2}h^{\delta\rho}\left(h_{\rho\beta,\alpha}+h_{\rho\alpha,\beta}- h_{\alpha\beta,\rho}\right)
\end{equation}
are the Christoffel symbols for the two-dimensional metric tensor $h$.
\item  $\rho=2$:
\begin{eqnarray*}
&&\frac{1}{2}G^{\delta 2}\left(G_{2\beta,\alpha}+G_{2\alpha,\beta}- G_{\alpha\beta,2}\right)\\
                             &=&\frac{\epsilon}{2}(-\varphi^{\delta})\left(\varphi_{\alpha,\beta}+\varphi_{\beta,\alpha}-[\epsilon^{-1}h_{\alpha\beta,2}+(\varphi_{\alpha}\varphi_{\beta})_{,2}]\right)\\
                             &=&\frac{-\epsilon}{2}\left\{\varphi^{\delta}\varphi_{\alpha,\beta}+\varphi^{\delta}\varphi_{\beta,\alpha}\right\}
\end{eqnarray*}
where the last line follows from the fact that
\begin{equation}
\epsilon^{-1}h_{\alpha\beta,2}+(\varphi_{\alpha}\varphi_{\beta})_{,2}=0,
\end{equation}
since the $\partial_{2}$ derivatives of $h_{\alpha\beta}$, and $\varphi_{\alpha}$ vanish.\\
\\
Clearly, the terms in curly brackets cancel in the sum of the $\rho\in\{0,1\}$ and $\rho=2$ cases above, and we have for $\delta, \alpha, \beta\in\{0,1\}$:
\begin{equation}
\Gamma^{\delta}_{\alpha\beta}=\gamma^{\delta}_{\alpha\beta}-\frac{\epsilon}{2}h^{\delta\zeta}\left(\varphi_{\beta}f_{\zeta\alpha}+\varphi_{\alpha}f_{\zeta\beta}\right),
\end{equation}
where,
\begin{equation}
f_{\alpha\beta}:=\varphi_{\beta,\alpha}-\varphi_{\alpha,\beta}.
\end{equation}
\end{itemize}

\item  $\Gamma^{2}_{\alpha\beta}$, $\alpha, \beta\in\{0,1\}$.
\begin{itemize}
\item  $\rho\in\{0,1\}$:
\begin{eqnarray*}
&&\frac{1}{2}G^{2 \rho}\left(G_{\rho\beta,\alpha}+G_{\rho\alpha,\beta}- G_{\alpha\beta,\rho}\right)\\
                             &=&\frac{1}{2}(-\varphi^{\rho})\left([h_{\rho\beta,\alpha}+\epsilon(\varphi_{\rho}\varphi_{\beta})_{,\alpha}]+[h_{\rho\alpha,\beta}+\epsilon(\varphi_{\rho}\varphi_{\alpha})_{,\beta}]- [h_{\alpha\beta,\rho}+\epsilon(\varphi_{\alpha}\varphi_{\beta})_{,\rho}]\right)\\
                             &=&\frac{1}{2}(-\varphi^{\rho})\left([h_{\rho\beta,\alpha}+\epsilon(\varphi_{\rho,\alpha}\varphi_{\beta}+\varphi_{\rho}\varphi_{\beta,\alpha})]+[h_{\rho\alpha,\beta}+\epsilon(\varphi_{\rho,\beta}\varphi_{\alpha}+\varphi_{\rho}\varphi_{\alpha,\beta})]\right)+\\
                             &-&\frac{1}{2}(-\varphi^{\rho})\left([h_{\alpha\beta,\rho}+\epsilon(\varphi_{\alpha,\rho}\varphi_{\beta}+\varphi_{\alpha}\varphi_{\beta,\rho})]\right)\\
                             &=&-\varphi_{\rho}\gamma^{\rho}_{\alpha\beta}-\frac{\epsilon}{2}\varphi^{\rho}\left((\varphi_{\rho,\alpha}\varphi_{\beta}+\varphi_{\rho}\varphi_{\beta,\alpha})+(\varphi_{\rho,\beta}\varphi_{\alpha}+\varphi_{\rho}\varphi_{\alpha,\beta})-(\varphi_{\alpha,\rho}\varphi_{\beta}+\varphi_{\alpha}\varphi_{\beta,\rho})\right)\\
                             &=&-\varphi_{\rho}\gamma^{\rho}_{\alpha\beta}+\frac{\epsilon}{2}\left(\varphi^{\rho}[\varphi_{\beta}f_{\rho\alpha}+\varphi_{\alpha}f_{\rho\beta})]-\left\{\varphi^{\rho}\varphi_{\rho}\varphi_{\alpha,\beta}+\varphi^{\rho}\varphi_{\rho}\varphi_{\beta,\alpha}\right\}\right)
\end{eqnarray*}
\item $\rho=2$:
\begin{eqnarray*}
&&\frac{1}{2}G^{2 2}\left(G_{2\beta,\alpha}+G_{2\alpha,\beta}- G_{\alpha\beta,2}\right)\\
                             &=&\frac{\epsilon}{2}(\epsilon^{-1}+\varphi^{\zeta}\varphi_{\zeta})\left(\varphi_{\alpha,\beta}+\varphi_{\beta,\alpha}\right),\,\,\text{since $G_{\alpha\beta,2}=0$,}\\
                             &=&\frac{1}{2}(\varphi_{\alpha,\beta}+\varphi_{\beta,\alpha})+\frac{\epsilon}{2}\left\{\varphi^{\zeta}\varphi_{\zeta}\varphi_{\alpha,\beta}+\varphi^{\zeta}\varphi_{\zeta}\varphi_{\beta,\alpha}\right\}
\end{eqnarray*}
Clearly, the terms in curly brackets cancel in the sum of the $\rho\in\{0,1\}$ and $\rho=2$ cases, and we have:
\begin{eqnarray*}
\Gamma^{2}_{\alpha\beta}&=&\frac{1}{2}(\varphi_{\alpha,\beta}+\varphi_{\beta,\alpha})-\varphi_{\zeta}\gamma^{\zeta}_{\alpha\beta}+\frac{\epsilon}{2}\varphi^{\zeta}[\varphi_{\beta}f_{\zeta\alpha}+\varphi_{\alpha}f_{\zeta\beta}]\\
                             &=&\frac{1}{2}(D_{\alpha}\varphi_{\beta}+D_{\beta}\varphi_{\alpha})+\frac{\epsilon}{2}\varphi^{\zeta}(\varphi_{\beta}f_{\zeta\alpha}+\varphi_{\alpha}f_{\zeta\beta}),
\end{eqnarray*}
where the last line follows from the fact that the Levi-Civita connection is symmetric (i.e. $\gamma^{\zeta}_{\alpha\beta}=\gamma^{\zeta}_{\beta\alpha}$), and
\begin{equation}
D_{\alpha}\varphi_{\beta}:=\varphi_{\beta,\alpha}-\varphi_{\zeta}\gamma^{\zeta}_{\alpha\beta}.
\end{equation}
\end{itemize}
\item  $\Gamma^{\delta}_{2\beta}$, $\delta, \beta\in\{0,1\}$.
\begin{itemize}
\item  $\rho\in\{0,1\}$:
\begin{eqnarray*}
&&\frac{1}{2}G^{\delta\rho}\left(G_{\rho\beta,2}+G_{\rho 2,\beta}- G_{2 \beta,\rho}\right)\\
                             &=&\frac{\epsilon}{2}h^{\delta\rho}\left(\varphi_{\rho,\beta}-\varphi_{\beta,\rho}\right),\,\,\text{since $G_{\rho\beta,2}=0$,}\\
                             &=&\frac{\epsilon}{2}h^{\delta\zeta}f_{\beta\zeta}.
\end{eqnarray*}
\item  $\rho=2$:  It is easy to see that
\begin{equation}
\frac{1}{2}G^{\delta 2}\left(G_{2\beta,2}+G_{2 2,\beta}- G_{2 \beta,2}\right)=0.
\end{equation}
This follows from observing that $G_{2 2,\beta}=\partial_{\beta}\epsilon=0$, and the $\partial_{2}$ derivative $G_{2\beta,2}$ vanishes.  Thus,
\begin{equation}
\Gamma^{\delta}_{2\beta}=\frac{\epsilon}{2}h^{\delta\zeta}f_{\beta\zeta}.
\end{equation}
\end{itemize}
\item  $\Gamma^{2}_{2\beta}$, $\beta\in\{0,1\}$.
\begin{itemize}
\item  $\rho\in\{0,1\}$:
\begin{eqnarray*}
&&\frac{1}{2}G^{\delta\rho}\left(G_{\rho\beta,2}+G_{\rho 2,\beta}- G_{2 \beta,\rho}\right)\\
                             &=&\frac{\epsilon}{2}(-\varphi^{\rho})\left(\varphi_{\rho,\beta}-\varphi_{\beta,\rho}\right),\,\,\text{since $G_{\rho\beta,2}=0$,}\\
                             &=&\frac{\epsilon}{2}\varphi^{\zeta}f_{\zeta\beta}.
\end{eqnarray*}
\item  $\rho=2$:  As above, it is not difficult to see that
\begin{equation}
\frac{1}{2}G^{\delta 2}\left(G_{2\beta,2}+G_{2 2,\beta}- G_{2 \beta,2}\right)=0.
\end{equation}
\end{itemize}
Thus,
\begin{equation}
\Gamma^{2}_{2\beta}=\frac{\epsilon}{2}\varphi^{\zeta}f_{\zeta\beta}.
\end{equation}
\item  $\Gamma^{\lambda}_{2 2}$, $\lambda\in\{0,1,2\}$.  It is also easy to see that
\begin{equation}
\frac{1}{2}G^{\lambda\rho}\left(G_{\rho 2,2}+G_{\rho 2,2}- G_{2 2,\rho}\right)=0.
\end{equation}
\end{enumerate}

\subsection{Spin Connection}\label{spin}
In this section we use our formulae for the Vielbein,
\begin{eqnarray}\label{vielform1}
E^{a}_{\alpha}&=&e^{a}_{\alpha},\,\,E^{2}_{2}=\sqrt{\epsilon},\,\,E^{2}_{\alpha}=\sqrt{\epsilon}\varphi_{\alpha},\,\,E^{a}_{2}=0\\\label{vielform2}
\widetilde{E}^{\alpha}_{a}&=&\widetilde{e}^{\alpha}_{a},\,\,\widetilde{E}^{2}_{2}=\frac{1}{\sqrt{\epsilon}},\,\,\widetilde{E}^{2}_{a}=-\varphi_{\zeta}\widetilde{e}^{\zeta}_{a},\,\,\widetilde{E}^{\alpha}_{2}=0,
\end{eqnarray}
\noindent
and our formulae for the Christoffel symbols for the Levi-Civita connection associated to our family of metrics $G:=g_{\epsilon}$ to compute the spin connections $[(A^{G})_{\mu}]^{A}_{B}:=[A_{\mu}]^{A}_{B}$ using Eq. \ref{spinviel},
\begin{equation}\label{spinsum}
[A_{\mu}]^{A}_{B}=E_{\nu}^{A}\,\widetilde{E}^{\lambda}_{B}\Gamma^{\nu}_{\mu\lambda}-\widetilde{E}^{\lambda}_{B}\partial_{\mu}E_{\lambda}^{A}.
\end{equation}
\noindent
We break this down into cases.
\begin{enumerate}\renewcommand{\theenumi}{\Roman{enumi}}
\item  $[A_{\alpha}]^{a}_{b}$, $a,b,\alpha\in\{0,1\}$.  Plug in the appropriate quantities from Eq.'s \ref{vielform1} and \ref{vielform2} and for the Christoffel symbols.
\begin{itemize}
\item  $\lambda, \nu\in\{0,1\}$:
\begin{eqnarray*}
E_{\nu}^{a}\,\widetilde{E}^{\lambda}_{b}\Gamma^{\nu}_{\alpha\lambda}-\widetilde{E}^{\lambda}_{b}\partial_{\alpha}E_{\lambda}^{a}&=&\widetilde{e}^{\zeta}_{b}e^{a}_{\delta}\left(\gamma^{\delta}_{\alpha\zeta}-\frac{\epsilon}{2}h^{\delta\rho}\left(\varphi_{\zeta}f_{\rho\alpha}+\varphi_{\alpha}f_{\rho\zeta}\right)\right)-\widetilde{e}^{\zeta}_{b}\partial_{\alpha}e^{a}_{\zeta}\\
                    &=&\widetilde{e}^{\zeta}_{b}\left[(e^{a}_{\delta}\gamma^{\delta}_{\alpha\zeta}-\partial_{\alpha}e^{a}_{\zeta})-\frac{\epsilon}{2}e^{a}_{\delta}h^{\delta\rho}\varphi_{\alpha}f_{\rho\zeta}\right]-\left\{\frac{\epsilon}{2}\widetilde{e}^{\zeta}_{b}e^{a}_{\delta}h^{\delta\rho}\varphi_{\zeta}f_{\rho\alpha}\right\}\\
                    &=&\widetilde{e}^{\zeta}_{b}\left[-D_{\alpha}e^{a}_{\zeta}-\frac{\epsilon}{2}e^{a}_{\delta}h^{\delta\rho}\varphi_{\alpha}f_{\rho\zeta}\right]-\left\{\frac{\epsilon}{2}\widetilde{e}^{\zeta}_{b}e^{a}_{\delta}h^{\delta\rho}\varphi_{\zeta}f_{\rho\alpha}\right\}
\end{eqnarray*}
\item  $\lambda=2$, $\nu\in\{0,1\}$:
\begin{eqnarray*}
E_{\nu}^{a}\,\widetilde{E}^{\lambda}_{b}\Gamma^{\nu}_{\alpha\lambda}-\widetilde{E}^{\lambda}_{b}\partial_{\alpha}E_{\lambda}^{a}&=&e^{a}_{\delta}(-\varphi_{\zeta}\widetilde{e}^{\zeta}_{b})\left(\frac{\epsilon}{2}\right)h^{\delta\rho}f_{\alpha\rho}\\
                    &=&\left\{\frac{\epsilon}{2}\widetilde{e}^{\zeta}_{b}e^{a}_{\delta}h^{\delta\rho}\varphi_{\zeta}f_{\rho\alpha}\right\}
\end{eqnarray*}

\end{itemize}
\noindent
These two cases, $\lambda, \nu\in\{0,1\}$ and $\lambda=2$, $\nu\in\{0,1\}$, are the only cases for which we get a non-zero contribution to $[A_{\alpha}]^{a}_{b}$, since the term $E_{\nu}^{a}\,\widetilde{E}^{\lambda}_{b}\Gamma^{\nu}_{\mu\lambda}$ always vanishes for $\nu=2$ when $a,b\in\{0,1\}$ by our formulae for the Vielbein, and the term $\widetilde{E}^{\lambda}_{b}\partial_{\mu}E_{\lambda}^{a}$ vanishes for $\lambda=2$ for the same reason.  After observing that the terms in the curly brackets from the cases, $\lambda, \nu\in\{0,1\}$ and $\lambda=2$, $\nu\in\{0,1\}$, cancel in our sum, we obtain:
\begin{equation}
[A_{\alpha}]^{a}_{b}=\widetilde{e}^{\zeta}_{b}\left[-D_{\alpha}e^{a}_{\zeta}-\frac{\epsilon}{2}e^{a}_{\delta}h^{\delta\rho}\varphi_{\alpha}f_{\rho\zeta}\right].
\end{equation}
\item  $[A_{\alpha}]^{a}_{2}$, $a,\alpha\in\{0,1\}$.  The only terms to contribute to the sum over $\lambda,\nu \in\{0,1,2\}$ in the sum of Eq. \ref{spinsum} are when $\lambda=2$ and $\nu\in\{0,1\}$, since $\widetilde{E}^{\lambda}_{2}=0$ for $\lambda\in\{0,1\}$.  Even when $\lambda=2$, the sum $\widetilde{E}^{\lambda}_{2}\partial_{\alpha}E_{\lambda}^{a}=0$ in Eq. \ref{spinsum} since $E_{2}^{a}=0$ for $a\in\{0,1\}$. Thus,
\begin{eqnarray*}
[A_{\alpha}]^{a}_{2}&=&\frac{1}{\sqrt{\epsilon}}e^{a}_{\delta}\Gamma^{\delta}_{\alpha 2}\\
                    &=&\frac{1}{\sqrt{\epsilon}}e^{a}_{\delta}\frac{\epsilon}{2}h^{\delta\zeta}f_{\alpha\zeta}\\
                    &=&\frac{\sqrt{\epsilon}}{2}e^{a}_{\delta}h^{\delta\zeta}f_{\alpha\zeta}
\end{eqnarray*}
By lowering the two dimensional index $[A_{\alpha}]_{a\,2}:=\widetilde{\eta}_{ab}[A_{\alpha}]^{b}_{2}$, with $\widetilde{\eta}_{ab}$ the two-dimensional Kronecker pairing, we obtain
\begin{equation}
[A_{\alpha}]_{a\,2}=-[A_{\alpha}]_{2\,a}=\widetilde{\eta}_{a\,b}\frac{\sqrt{\epsilon}}{2}e^{b}_{\delta}h^{\delta\zeta}f_{\alpha\zeta}.
\end{equation}
\item  $[A_{2}]^{a}_{b}$, $a,b\in\{0,1\}$.  The only terms to contribute to the sum over $\lambda,\nu \in\{0,1,2\}$ in the sum of Eq. \ref{spinsum} are when $\lambda,\nu \in\{0,1\}$.  All other terms vanish by our formulae for the Vielbein and the Christoffel symbols.  Note that the sum $\widetilde{E}^{\lambda}_{b}\partial_{2}E_{\lambda}^{a}=0$ in Eq. \ref{spinsum} since all $\partial_{2}$ derivatives vanish for the Vielbein.  Thus,
\begin{eqnarray*}
[A_{2}]^{a}_{b}&=&\widetilde{e}^{\zeta}_{b}e^{a}_{\delta}\Gamma^{\delta}_{\zeta 2}\\
                    &=&\widetilde{e}^{\zeta}_{b}e^{a}_{\delta}\frac{\epsilon}{2}h^{\delta\rho}f_{\zeta\rho}\\
                    &=&\frac{\epsilon}{2}\widetilde{e}^{\zeta}_{b}e^{a}_{\delta}h^{\delta\rho}f_{\zeta\rho}
\end{eqnarray*}
\item  $[A_{2}]^{a}_{2}$, $a,\alpha\in\{0,1\}$.  First, the sum $\widetilde{E}^{\lambda}_{2}\partial_{2}E_{\lambda}^{a}=0$ in Eq. \ref{spinsum} since all $\partial_{2}$ derivatives vanish for the Vielbein.  The only term to contribute to the sum over $\lambda,\nu \in\{0,1,2\}$ in the sum of Eq. \ref{spinsum} is $\nu \in\{0,1\}$ and $\lambda=2$.  Thus,
\begin{eqnarray*}
[A_{2}]^{a}_{2}&=&\widetilde{E}^{2}_{2}e^{a}_{\delta}\Gamma^{\delta}_{2 2}\\
               &=&0
\end{eqnarray*}
since $\Gamma^{\delta}_{2 2}=0$.  By lowering the two dimensional index $[A_{2}]_{a\,2}:=\widetilde{\eta}_{ab}[A_{\alpha}]^{b}_{2}$, and using our anti-symmetry properties, we see
\begin{equation}
[A_{2}]_{2\,a}=[A_{2}]_{a\,2}=0.
\end{equation}
\item  $[A_{\mu}]^{2}_{2}$, $\mu\in\{0,1,2\}$.  Lastly, since $[A_{\mu}]_{a\,b}$ is anti-symmetric in $a,b$, we see trivially that
\end{enumerate}
\begin{equation}
[A_{\mu}]^{2}_{2}=0.
\end{equation}
for all $\mu\in\{0,1,2\}$.

\subsection{Reduced Spin Connection}\label{spinred}
In this section we compute the corresponding vector valued one-forms $A^{C}_{\mu}$ for our spin connections $[(A)_{\mu}]^{A}_{B}$ defined by the relation
\begin{equation}
[(A)_{\mu}]_{A\,B}:=\eta_{AC}[A_{\mu}]^{C}_{B}=\epsilon_{ABC}A^{C}_{\mu}.
\end{equation}
Contracting with the Levi-Civita symbol $\epsilon^{ABC}$, we have
\begin{equation}\label{redspin}
A^{C}_{\mu}=\frac{1}{2}\epsilon^{ABC}\eta_{AD}[A_{\mu}]^{D}_{B}.
\end{equation}
Eq. \ref{redspin} combined with our formulae for the spin connections $[A_{\mu}]^{D}_{B}$ are the main relations that we use to compute $A^{C}_{\mu}$ throughout this section.  As usual, we do this computation in cases.
\begin{enumerate}\renewcommand{\theenumi}{\Roman{enumi}}
\item  $A^{2}_{\alpha}$, $\alpha\in\{0,1\}$.  Let $\widetilde{\eta}_{ab}$ denote the two-dimensional Kronecker pairing as usual.  Eq. \ref{redspin} gives us
\begin{eqnarray}
A^{2}_{\alpha}&=&\frac{1}{2}\epsilon^{AB2}\eta_{AD}[A_{\alpha}]^{D}_{B}\\
              &=&\frac{1}{2}\left[\widetilde{\eta}_{0a}[A_{\alpha}]^{a}_{1}-\widetilde{\eta}_{1a}[A_{\alpha}]^{a}_{0}\right]\\
              &=&\frac{1}{2}\left[\widetilde{\eta}_{0a}\widetilde{e}^{\zeta}_{1}\left[-D_{\alpha}e^{a}_{\zeta}-\frac{\epsilon}{2}e^{a}_{\delta}h^{\delta\rho}\varphi_{\alpha}f_{\rho\zeta}\right]-\widetilde{\eta}_{1a}\widetilde{e}^{\zeta}_{0}\left[-D_{\alpha}e^{a}_{\zeta}-\frac{\epsilon}{2}e^{a}_{\delta}h^{\delta\rho}\varphi_{\alpha}f_{\rho\zeta}\right]\right]\\\label{curlycomp}
              &=&\left\{-\frac{1}{2}(\widetilde{\eta}_{0a}\widetilde{e}^{\zeta}_{1}-\widetilde{\eta}_{1a}\widetilde{e}^{\zeta}_{0})D_{\alpha}e^{a}_{\zeta}\right\}+\left\{-\frac{\epsilon}{4}e^{a}_{\delta}h^{\delta\rho}\varphi_{\alpha}f_{\rho\zeta}(\widetilde{\eta}_{0a}\widetilde{e}^{\zeta}_{1}-\widetilde{\eta}_{1a}\widetilde{e}^{\zeta}_{0})\right\}
\end{eqnarray}
We compute the quantities in the curly brackets of Eq. \ref{curlycomp} separately.  First we recall that the spin connection $(\omega_{\alpha})^{a}_{b}$ on $\Sigma$ is defined by:
\begin{eqnarray}
(\omega_{\alpha})^{a}_{b}&:=&\widetilde{e}^{\zeta}_{b}\partial_{\alpha}e^{a}_{\zeta}-e^{a}_{\delta}\widetilde{e}^{\zeta}_{b}\gamma^{\delta}_{\alpha \zeta}\\\label{spinformu}
                         &=&\widetilde{e}^{\zeta}_{b}D_{\alpha}e^{a}_{\zeta}
\end{eqnarray}
Then $\omega_{\alpha}$ is defined by the relation
\begin{equation}\label{twospin}
\omega_{\alpha,ab}=\epsilon_{ab}\omega_{\alpha}=\widetilde{\eta}_{ac}(\omega_{\alpha})^{c}_{b}
\end{equation}
Thus, we compute the first term in Eq. \ref{curlycomp}:
\begin{eqnarray*}
\left\{-\frac{1}{2}(\widetilde{\eta}_{0a}\widetilde{e}^{\zeta}_{1}-\widetilde{\eta}_{1a}\widetilde{e}^{\zeta}_{0})D_{\alpha}e^{a}_{\zeta}\right\}&=&-\frac{1}{2}(\omega_{\alpha,01}-\omega_{\alpha,10}),\,\,\text{by Eq.'s \ref{spinformu} and \ref{twospin},}\\
                                                                                                                                                 &=&-\frac{1}{2}(2\omega_{\alpha,01}),\,\,\text{by anti-symmetry of $\omega_{\alpha,ab}$,}\\
                                                                                                                                                 &=&-\omega_{\alpha},\,\,\text{by Eq. \ref{twospin}.}
\end{eqnarray*}
Before computing the second term in Eq. \ref{curlycomp}, we recall the following relations:
\begin{equation}\label{zweirel}
\widetilde{e}^{\zeta}_{q}=h^{\zeta\lambda}\widetilde{\eta}_{qb}e^{b}_{\lambda},
\end{equation}
\begin{equation}\label{frel}
f_{\rho\zeta}=\sqrt{h}f\epsilon_{\rho\zeta},
\end{equation}
\begin{equation}\label{eprel}
\epsilon_{ab}=\widetilde{\eta}_{0a}\widetilde{\eta}_{1b}-\widetilde{\eta}_{1a}\widetilde{\eta}_{0b},
\end{equation}
\begin{equation}\label{detrel}
\epsilon^{\delta\lambda}\epsilon_{ab}e^{a}_{\delta}e^{b}_{\lambda}=2\sqrt{h}.
\end{equation}
where
\begin{equation}\label{detrel2}
\epsilon^{\delta\lambda}=|h|h^{\rho\delta}h^{\zeta\lambda}\epsilon_{\rho\zeta}
\end{equation}
in Eq. \ref{detrel}.  Using Eq.'s \ref{zweirel} and \ref{frel}, we compute the second term in curly brackets from Eq. \ref{curlycomp}:
\begin{eqnarray*}
\left\{-\frac{\epsilon}{4}e^{a}_{\delta}h^{\delta\rho}\varphi_{\alpha}f_{\rho\zeta}(\widetilde{\eta}_{0a}\widetilde{e}^{\zeta}_{1}-\widetilde{\eta}_{1a}\widetilde{e}^{\zeta}_{0})\right\}
&=&-\frac{\epsilon}{4}\sqrt{h}f\varphi_{\alpha}(e^{a}_{\delta}e^{b}_{\lambda})\left(h^{\rho\delta}h^{\zeta\lambda}\epsilon_{\rho\zeta}\right)\left(\widetilde{\eta}_{0a}\widetilde{\eta}_{1b}-\widetilde{\eta}_{1a}\widetilde{\eta}_{0b} \right)\\
&=&-\frac{\epsilon}{4}(\sqrt{h})^{-1}f\varphi_{\alpha}\left(e^{a}_{\delta}e^{b}_{\lambda}\epsilon^{\delta\lambda}\epsilon_{ab}\right),\,\,\text{by Eq.'s \ref{eprel} and \ref{detrel2},}\\
&=&-\frac{\epsilon}{2}(\sqrt{h})^{-1}f\varphi_{\alpha}\sqrt{h},\,\,\text{by Eq. \ref{detrel},}\\
&=&-\frac{\epsilon}{2}f\varphi_{\alpha}
\end{eqnarray*}
Thus, we have:
\begin{equation}
A^{2}_{\alpha}=-\omega_{\alpha}-\frac{\epsilon}{2}f\varphi_{\alpha}.
\end{equation}

\item $A^{2}_{2}$.  As before, we compute:
\begin{eqnarray*}
A^{2}_{2}&=&\frac{1}{2}\epsilon^{AB2}\eta_{AD}[A_{2}]^{D}_{B}\\
              &=&\frac{1}{2}\left[\widetilde{\eta}_{0a}[A_{2}]^{a}_{1}-\widetilde{\eta}_{1a}[A_{2}]^{a}_{0}\right]\\
              &=&\frac{1}{2}\left[\widetilde{\eta}_{0a}\left[\frac{\epsilon}{2}\widetilde{e}^{\zeta}_{1}e^{a}_{\delta}h^{\delta\rho}f_{\zeta\rho}\right]-\widetilde{\eta}_{1a}\left[\frac{\epsilon}{2}\widetilde{e}^{\zeta}_{0}e^{a}_{\delta}h^{\delta\rho}f_{\zeta\rho}\right]\right]\\
              &=&-\frac{\epsilon}{4}\sqrt{h}f e^{b}_{\lambda}e^{a}_{\delta}(h^{\delta\rho}h^{\zeta\lambda}\epsilon_{\rho\zeta})\left(\widetilde{\eta}_{0a}\widetilde{\eta}_{1b}-\widetilde{\eta}_{1a}\widetilde{\eta}_{0b} \right),\,\,\text{by Eq.'s \ref{zweirel} and \ref{frel},}\\
              &=&-\frac{\epsilon}{4}(\sqrt{h})^{-1}f (e^{a}_{\delta}e^{b}_{\lambda}\epsilon^{\delta\lambda}\epsilon_{ab}),\,\,\text{by Eq.'s \ref{eprel} and \ref{detrel2},}\\
              &=&-\frac{\epsilon}{2}(\sqrt{h})^{-1}f \sqrt{h},\,\,\text{by Eq. \ref{detrel},}\\
              &=&-\frac{\epsilon}{2}f
\end{eqnarray*}
Thus, we have:
\begin{equation}
A^{2}_{2}=-\frac{\epsilon}{2}f.
\end{equation}

\item $A^{a}_{\alpha}$, $a,\alpha\in\{0,1\}$.  Before we perform this computation we recall:
\begin{equation}\label{zw2}
\widetilde{e}^{\delta}_{a}=(\sqrt{h})^{-1}\epsilon^{\lambda\delta}\epsilon_{ab}e^{b}_{\lambda}
\end{equation}

Let $\widehat{a}\in\{0,1\}\backslash\{a\}$ be the element of $\{0,1\}$ that represents the compliment of $a\in\{0,1\}$.  We then compute:
\begin{eqnarray*}
A^{a}_{\alpha}&=&\frac{1}{2}\epsilon^{ABa}\eta_{AD}[A_{\alpha}]^{D}_{B}\\
              &=&\frac{1}{2}\epsilon_{a\widehat{a}}\left[\eta_{\widehat{a}D}[A_{\alpha}]^{D}_{2}-\eta_{2D}[A_{\alpha}]^{D}_{\widehat{a}}\right]\\
              &=&\frac{1}{2}\epsilon_{a\widehat{a}}\left[[A_{\alpha}]_{\widehat{a}2}-[A_{\alpha}]_{2\widehat{a}}\right]\\
              &=&\epsilon_{a\widehat{a}}[A_{\alpha}]_{\widehat{a}2},\,\,\text{by anti-symmetry of $[A_{\alpha}]_{ab}$,}\\
              &=&\epsilon_{a\widehat{a}}\widetilde{\eta}_{\widehat{a}\,b}\frac{\sqrt{\epsilon}}{2}e^{b}_{\delta}h^{\delta\zeta}f_{\alpha\zeta}\\
              &=&\epsilon_{a\widehat{a}}\frac{\sqrt{\epsilon}}{2}\sqrt{h}\,f\,\widetilde{e}^{\zeta}_{\widehat{a}}\,\epsilon_{\alpha\zeta},\,\,\text{by Eq.'s \ref{zweirel} and \ref{frel},}\\
              &=&\epsilon_{a\widehat{a}}\frac{\sqrt{\epsilon}}{2}\sqrt{h}\,f\,[(\sqrt{h})^{-1}\epsilon^{\lambda\zeta}\epsilon_{\widehat{a}b}e^{b}_{\lambda}]\,\epsilon_{\alpha\zeta},\,\,\text{by Eq. \ref{zw2},}\\
              &=&\frac{\sqrt{\epsilon}}{2}\,f\,[\epsilon^{\lambda\zeta}\,\epsilon_{\alpha\zeta}\epsilon_{a\widehat{a}}\epsilon_{\widehat{a}b}e^{b}_{\lambda}]\\
              &=&\frac{\sqrt{\epsilon}}{2}\,f\,[\delta^{\lambda}_{\alpha}e^{a}_{\lambda}]\\
              &=&\frac{\sqrt{\epsilon}}{2}\,f\,e^{a}_{\alpha}.
\end{eqnarray*}
Thus, we have:
\begin{equation}
A^{a}_{\alpha}=\frac{\sqrt{\epsilon}}{2}\,f\,e^{a}_{\alpha}.
\end{equation}
\item  $A^{a}_{2}$, $a\in\{0,1\}$.
\begin{eqnarray*}
A^{a}_{2}&=&\frac{1}{2}\epsilon^{ABa}\eta_{AD}[A_{2}]^{D}_{B}\\
              &=&\frac{1}{2}\epsilon_{a\widehat{a}}\left[\eta_{\widehat{a}D}[A_{2}]^{D}_{2}-\eta_{2D}[A_{2}]^{D}_{\widehat{a}}\right]\\
              &=&\frac{1}{2}\epsilon_{a\widehat{a}}\left[[A_{2}]_{\widehat{a}2}-[A_{2}]_{2\widehat{a}}\right]\\
              &=&\epsilon_{a\widehat{a}}[A_{2}]_{\widehat{a}2},\,\,\text{by anti-symmetry of $[A_{\alpha}]_{ab}$,}\\
              &=&0,\,\,\text{since $[A_{2}]_{a2}=0$ in general.}
\end{eqnarray*}
Thus,
\begin{equation}
A^{a}_{2}=0.
\end{equation}
\end{enumerate}

\subsection{Reduced Gravitational Chern-Simons}\label{reduced}
In this section we compute the gravitational Chern-Simons term $CS(A^{g_{\epsilon}})$ in terms of reduced quantities using Eq. \ref{compeq2}:
\begin{equation}\label{redcomp}
CS(A^{G})=-\frac{1}{2\pi}\int_{X}d^{3}x\,\,\epsilon^{\mu\nu\lambda}\,\,\left(\eta_{AB}A^{A}_{\mu}\partial_{\nu} A^{B}_{\lambda}\right)+\frac{1}{\pi}\int_{X}d^{3}x\,\,\text{det}A^{C}_{\mu}.
\end{equation}
\begin{enumerate}\renewcommand{\theenumi}{\Roman{enumi}}
\item  We first compute $\epsilon^{\mu\nu\lambda}\eta_{AB}A^{A}_{\mu}\partial_{\nu} A^{B}_{\lambda}$ from the first integral term.  First observe that $A^{a}_{\mu}\partial_{\nu} A^{a}_{\lambda}=0$ for any $a\in\{0,1\}$ and any permutation $\sigma(012)=\mu\nu\lambda$, since if $\nu=2$, then all $\partial_{2}$ derivatives vanish, and if $\nu\neq 2$ then $A^{a}_{2}=0$ by our previous results.  Thus, we only need compute the term $\epsilon^{\mu\nu\lambda}A^{2}_{\mu}\partial_{\nu} A^{2}_{\lambda}$, where $\nu\neq 2$.  We do this in cases.

\begin{itemize}
\item  $(\mu,\nu,\lambda)=(2,0,1)$:
\begin{eqnarray*}
\epsilon^{201}A^{2}_{2}\partial_{0} A^{2}_{1}&=&[-\frac{\epsilon}{2}f]\cdot \partial_{0}[-\omega_{1}-\frac{\epsilon}{2}f\varphi_{1}]\\
                                             &=&[\frac{\epsilon}{2}f]\cdot \partial_{0}[\omega_{1}+\frac{\epsilon}{2}f\varphi_{1}]
\end{eqnarray*}
\item $(\mu,\nu,\lambda)=(2,1,0)$:
\begin{eqnarray*}
\epsilon^{210}A^{2}_{2}\partial_{1} A^{2}_{0}&=&-[-\frac{\epsilon}{2}f]\cdot \partial_{1}[-\omega_{0}-\frac{\epsilon}{2}f\varphi_{0}]\\
                                             &=&-[\frac{\epsilon}{2}f]\cdot \partial_{1}[\omega_{0}+\frac{\epsilon}{2}f\varphi_{0}]
\end{eqnarray*}
\item $(\mu,\nu,\lambda)=(0,1,2)$:
\begin{eqnarray*}
\epsilon^{012}A^{2}_{0}\partial_{1} A^{2}_{2}&=&[-\omega_{0}-\frac{\epsilon}{2}f\varphi_{0}]\cdot \partial_{1}[-\frac{\epsilon}{2}f]\\
                                             &=&[\omega_{0}+\frac{\epsilon}{2}f\varphi_{0}]\cdot \partial_{1}[\frac{\epsilon}{2}f]
\end{eqnarray*}
\item $(\mu,\nu,\lambda)=(1,0,2)$:
\begin{eqnarray*}
\epsilon^{102}A^{2}_{1}\partial_{0} A^{2}_{2}&=&-[-\omega_{1}-\frac{\epsilon}{2}f\varphi_{1}]\cdot \partial_{0}[-\frac{\epsilon}{2}f]\\
                                             &=&-[\omega_{1}+\frac{\epsilon}{2}f\varphi_{1}]\cdot \partial_{0}[\frac{\epsilon}{2}f]
\end{eqnarray*}
\end{itemize}
Adding these four cases together and grouping terms by powers of $\epsilon$ we obtain:
\begin{eqnarray*}
\epsilon^{\mu\nu\lambda}A^{2}_{\mu}\partial_{\nu} A^{2}_{\lambda}&=&\left(\frac{\epsilon}{2}\right)[f(\partial_{0}\omega_{1}-\partial_{0}\omega_{1})+(\omega_{0}\partial_{1}f-\omega_{1}\partial_{0}f)]+\\
               &+& \left(\frac{\epsilon}{2}\right)^{2}[(f\partial_{0}(f\varphi_{1})-f\partial_{1}(f\varphi_{0}))+(f\varphi_{0}\partial_{1}f-f\varphi_{1}\partial_{0}f)]\\
               &=&\left(\frac{\epsilon}{2}\right)[2f(\partial_{0}\omega_{1}-\partial_{1}\omega_{0})+\{\partial_{1}(\omega_{0}f)-\partial_{0}(\omega_{1}f)\}]+\\
               &+& \left(\frac{\epsilon}{2}\right)^{2}[f^{2}(\partial_{0}\varphi_{1}-\partial_{1}\varphi_{0})]\\
\end{eqnarray*}
The term in curly brackets in the second last line above yields a global exact form on $\Sigma$, and since $\partial\Sigma=\emptyset$, Stokes' theorem implies that this term vanishes in the integral of Eq. \ref{redcomp}.  It is also well known that the term in the second last line above, $\partial_{0}\omega_{1}-\partial_{1}\omega_{0}=\frac{1}{2}\sqrt{h}r$, where $r\in\Omega^{0}_{orb}(\Sigma)$ is the (orbifold) scalar curvature of $(\Sigma, h)$. Also, the term $\partial_{0}\varphi_{1}-\partial_{1}\varphi_{0}=f_{01}=\sqrt{h}f$.  Thus, we may write:
\begin{equation}\label{m1}
\int_{X}d^{3}x\,\,\epsilon^{\mu\nu\lambda}\,\,\left(\eta_{AB}A^{A}_{\mu}\partial_{\nu} A^{B}_{\lambda}\right)=\int_{S^{1}}dx^{2}\int_{\Sigma}dx^{0}\wedge dx^{1}\sqrt{h}\left[\left(\frac{\epsilon}{2}\right) f r+\left(\frac{\epsilon}{2}\right)^{2}f^{3}\right]
\end{equation}
This completes our computation of the first integral term in Eq. \ref{redcomp}.
\item  We now compute the second integral term $\text{det}A^{C}_{\mu}$ from Eq. \ref{redcomp}. For this we note that $\text{det}A^{C}_{\mu}$ may be computed directly, since:
\begin{equation}
A^{C}_{\mu}=\begin{bmatrix}
\frac{\sqrt{\epsilon}}{2}e^{a}_{\alpha}f & 0 \\
-\omega_{\alpha}-\frac{\epsilon}{2}f\varphi_{\alpha}  & -\frac{\epsilon}{2}f
\end{bmatrix}.
\end{equation}
Thus,
\begin{eqnarray*}
\text{det}A^{C}_{\mu}&=&\text{det}\left(\frac{\sqrt{\epsilon}}{2}e^{a}_{\alpha}f\right)\cdot \left(-\frac{\epsilon}{2}f\right)\\
                     &=&\left(\frac{\epsilon}{4}\,f^{2}\,\text{det}(e^{a}_{\alpha})\right)\cdot \left(-\frac{\epsilon}{2}f\right)\\
                     &=&-\frac{1}{2}\left(\frac{\epsilon}{2}\right)^{2}\sqrt{h}f^{3},\,\,\text{since $\text{det}(e^{a}_{\alpha})=\sqrt{h}$}.
\end{eqnarray*}
Thus,
\begin{equation}\label{m2}
\int_{X}d^{3}x\,\,\text{det}A^{C}_{\mu}=\int_{S^{1}}dx^{2}\int_{\Sigma}dx^{0}\wedge dx^{1}\sqrt{h}\left[-\frac{1}{2}\left(\frac{\epsilon}{2}\right)^{2}f^{3}\right].
\end{equation}

\end{enumerate}
Adding our main results from Eq.'s \ref{m1} and \ref{m2}, we obtain:
\begin{equation}
CS(A^{g_{\epsilon}})=-\frac{1}{4\pi}\int_{S^{1}}dx^{2}\int_{\Sigma}dx^{0}\wedge dx^{1}\sqrt{h}(\epsilon f r+\epsilon^{2} f^{3}).
\end{equation}

\bibliography{GSbiblio}

\begin{thebibliography}{14}
\expandafter\ifx\csname natexlab\endcsname\relax\def\natexlab#1{#1}\fi
\expandafter\ifx\csname bibnamefont\endcsname\relax
  \def\bibnamefont#1{#1}\fi
\expandafter\ifx\csname bibfnamefont\endcsname\relax
  \def\bibfnamefont#1{#1}\fi
\expandafter\ifx\csname citenamefont\endcsname\relax
  \def\citenamefont#1{#1}\fi
\expandafter\ifx\csname url\endcsname\relax
  \def\url#1{\texttt{#1}}\fi
\expandafter\ifx\csname urlprefix\endcsname\relax\def\urlprefix{URL }\fi
\providecommand{\bibinfo}[2]{#2}
\providecommand{\eprint}[2][]{\url{#2}}

\bibitem[{\citenamefont{Guralnik et~al.}(2003)\citenamefont{Guralnik, Iorio,
  Jackiw, and Pi}}]{gijp}
\bibinfo{author}{\bibfnamefont{G.}~\bibnamefont{Guralnik}},
  \bibinfo{author}{\bibfnamefont{A.}~\bibnamefont{Iorio}},
  \bibinfo{author}{\bibfnamefont{R.}~\bibnamefont{Jackiw}}, \bibnamefont{and}
  \bibinfo{author}{\bibfnamefont{S.-Y.} \bibnamefont{Pi}},
  \bibinfo{journal}{Ann. Physics} \textbf{\bibinfo{volume}{308}},
  \bibinfo{pages}{222} (\bibinfo{year}{2003}).

\bibitem[{\citenamefont{Deser et~al.}(1982)\citenamefont{Deser, Jackiw, and
  Templeton}}]{djt}
\bibinfo{author}{\bibfnamefont{S.}~\bibnamefont{Deser}},
  \bibinfo{author}{\bibfnamefont{R.}~\bibnamefont{Jackiw}}, \bibnamefont{and}
  \bibinfo{author}{\bibfnamefont{S.}~\bibnamefont{Templeton}},
  \bibinfo{journal}{Annals Phys.} \textbf{\bibinfo{volume}{140}},
  \bibinfo{pages}{372} (\bibinfo{year}{1982}).

\bibitem[{\citenamefont{Biquard et~al.}(2007)\citenamefont{Biquard, Herzlich,
  and Rumin}}]{bhr}
\bibinfo{author}{\bibfnamefont{O.}~\bibnamefont{Biquard}},
  \bibinfo{author}{\bibfnamefont{M.}~\bibnamefont{Herzlich}}, \bibnamefont{and}
  \bibinfo{author}{\bibfnamefont{M.}~\bibnamefont{Rumin}},
  \bibinfo{journal}{Ann. Scient. Ec. Norm. Sup.} \textbf{\bibinfo{volume}{40}},
  \bibinfo{pages}{589} (\bibinfo{year}{2007}).

\bibitem[{\citenamefont{Bismut and Cheeger}(1989)}]{bc}
\bibinfo{author}{\bibfnamefont{J.}~\bibnamefont{Bismut}} \bibnamefont{and}
  \bibinfo{author}{\bibfnamefont{J.}~\bibnamefont{Cheeger}},
  \bibinfo{journal}{J. Amer. Math. Soc.} \textbf{\bibinfo{volume}{2}},
  \bibinfo{pages}{33} (\bibinfo{year}{1989}).

\bibitem[{\citenamefont{Dai}(1991)}]{dai}
\bibinfo{author}{\bibfnamefont{X.}~\bibnamefont{Dai}}, \bibinfo{journal}{J.
  Amer. Math. Soc.} \textbf{\bibinfo{volume}{4}}, \bibinfo{pages}{265}
  (\bibinfo{year}{1991}).

\bibitem[{\citenamefont{Jeffrey and McLellan}(2010)}]{jm2}
\bibinfo{author}{\bibfnamefont{L.~C.} \bibnamefont{Jeffrey}} \bibnamefont{and}
  \bibinfo{author}{\bibfnamefont{B.~D.~K.} \bibnamefont{McLellan}}, in
  \emph{\bibinfo{booktitle}{Chern-Simons Gauge Theory, 20 Years After, Conf.
  Proc.}} (\bibinfo{year}{2010}).

\bibitem[{\citenamefont{Witten}(1989)}]{w3}
\bibinfo{author}{\bibfnamefont{E.}~\bibnamefont{Witten}},
  \bibinfo{journal}{Commun. Math. Phys.} \textbf{\bibinfo{volume}{121}},
  \bibinfo{pages}{351} (\bibinfo{year}{1989}).

\bibitem[{\citenamefont{Atiyah}(1990)}]{at}
\bibinfo{author}{\bibfnamefont{M.~F.} \bibnamefont{Atiyah}},
  \bibinfo{journal}{Topology} \textbf{\bibinfo{volume}{29}}, \bibinfo{pages}{1}
  (\bibinfo{year}{1990}).

\bibitem[{\citenamefont{Boyer and Galicki}(2008)}]{bg}
\bibinfo{author}{\bibfnamefont{C.~P.} \bibnamefont{Boyer}} \bibnamefont{and}
  \bibinfo{author}{\bibfnamefont{K.}~\bibnamefont{Galicki}},
  \emph{\bibinfo{title}{Sasakian Geometry}} (\bibinfo{publisher}{Oxford
  University Press}, \bibinfo{year}{2008}).

\bibitem[{\citenamefont{Itoh}(1997)}]{itoh}
\bibinfo{author}{\bibfnamefont{M.}~\bibnamefont{Itoh}}, \bibinfo{journal}{Proc.
  Japan Acad.} \textbf{\bibinfo{volume}{72}}, \bibinfo{pages}{58}
  (\bibinfo{year}{1997}).

\bibitem[{\citenamefont{Blair}(1976)}]{b}
\bibinfo{author}{\bibfnamefont{D.~E.} \bibnamefont{Blair}},
  \emph{\bibinfo{title}{Riemannian Geometry of Contact and Symplectic
  Manifolds}}, vol. \bibinfo{volume}{509} of \emph{\bibinfo{series}{Lecture
  notes in mathematics}} (\bibinfo{publisher}{Springer-Verlag, Berlin},
  \bibinfo{year}{1976}).

\bibitem[{\citenamefont{Bergmann}(1947)}]{berg}
\bibinfo{author}{\bibfnamefont{P.}~\bibnamefont{Bergmann}},
  \emph{\bibinfo{title}{Introduction to the theory of relativity}}
  (\bibinfo{publisher}{Prentice Hall, New York, NY}, \bibinfo{year}{1947}),
  chap.~\bibinfo{chapter}{17}.

\bibitem[{\citenamefont{Satake}(1957)}]{sat}
\bibinfo{author}{\bibfnamefont{I.}~\bibnamefont{Satake}},
  \bibinfo{journal}{Journal of the Mathematical Society of Japan}
  \textbf{\bibinfo{volume}{9}}, \bibinfo{pages}{464} (\bibinfo{year}{1957}).

\bibitem[{\citenamefont{Nicolaescu}(2000)}]{nic}
\bibinfo{author}{\bibfnamefont{L.~I.} \bibnamefont{Nicolaescu}},
  \bibinfo{journal}{Comm. Anal. Geo.} \textbf{\bibinfo{volume}{8}},
  \bibinfo{pages}{1027} (\bibinfo{year}{2000}).

\end{thebibliography}

\end{document}